\documentclass[authoryear,12pt]{elsarticle}
\usepackage{graphicx}
\usepackage{amsfonts}
\usepackage{amssymb}
\usepackage{amsmath}
\usepackage{threeparttable}
\usepackage{multirow}
\usepackage{subfig}
\usepackage[colorlinks]{hyperref}
\usepackage[scale=0.85]{geometry}

\usepackage{color}
\usepackage{mathtools} 

\begin{document}

\begin{frontmatter}
	
	\title{On the closure requirement for VOF simulations with RANS modeling}
	\author[a]{Wenyuan~Fan\corref{cor1}}
	\ead{wf@kth.se}
	\author[a,b]{Henryk~Anglart}
	\cortext[cor1]{Corresponding author, Tel: +46 737652495.}
	\address[a]{Nuclear Engineering Division, Department of Physics, KTH Royal Institute of Technology, 106 91 Stockholm, Sweden}
	\address[b]{Institute of Heat Engineering, Warsaw University of Technology, 21/25 Nowowiejska Street, 00-665 Warsaw, Poland}
	
	\begin{abstract}

	The volume of fluid (VOF) method is increasingly used in computational fluid dynamics (CFD) simulations of turbulent two-phase flows. The Reynolds-Averaged Navier-Stokes (RANS) approach is an economic and practical way for turbulent VOF simulations. Even though RANS-VOF simulations are widely conducted, the underlying physics is barely discussed. This study reveals the very basic closure requirement for RANS-VOF simulations.

	\end{abstract}
	
\end{frontmatter}

	\section{Introduction}
	
	The volume of fluid (VOF) method \citep{Hirt1981} is widely used in computational fluid dynamics (CFD) simulations of two-phase flows. An overview of the VOF method will be given in Section \ref{sect:vof}. Even though the traditional VOF-related topics, e.g., interface reconstruction and surface tension calculation, are still extensively studied, another topic, namely turbulence modeling for VOF, is becoming more and more popular. This popularity is undoubtedly demand-oriented since turbulent two-phase flows are ubiquitous in nature and industrial applications. Similarly to single-phase CFD, the turbulence modeling for VOF can also be categorized into three main groups: direct numerical simulations (DNSs) \citep{Ling2017,Ling2019}, large eddy simulations (LESs) \citep{Liovic2007,Liovic2012,Saxena2019} and Reynolds-Averaged Navier-Stokes (RANS) simulations \citep{Fan2018j,Saxena2019}.
	
	This study focuses on RANS modeling for VOF due to following two reasons. Firstly and unsurprisingly, DNS and LES are still too computationally expensive to be used in most practical applications. On the other hand, even though widely used, RANS modeling for VOF is the least developed one when coming to the underlying physics. Starting with fundamental concepts in VOF and RANS modeling, this study reveals the very basic closure assumption that has been implicitly made in common CFD software for RANS-VOF simulations.

	\section{Overview of the VOF method}
	\label{sect:vof}
	
	The core concept of the VOF method is to introduce the volumetric fraction (of the primary phase), $\alpha$, to model an immiscible two-phase system. Then the two-phase system could be treated as a single-phase system by introducing the mixture density:
	\begin{equation}
	\label{eq:average density}
	\rho = \alpha \rho _1+(1-\alpha)\rho _2,
	\end{equation}
	and the mixture viscosity:
	\begin{equation}
	\label{eq:average viscosity}
	\mu = \alpha \mu _1+(1-\alpha)\mu _2,
	\end{equation}
	where $\rho _1$ and $\rho _2$ denote the density for the primary and the secondary phase, respectively; $\mu _1$ and $\mu _2$ are the dynamic viscosity for the primary and the secondary phase, respectively. Using these two properties of the mixture, the two-phase system could be described by the following governing equations. 

    \subsection{Incompressible flow condition}
    
    For an isothermal two-phase system with two immiscible incompressible fluids, the flow field of each fluid is obviously divergence-free, therefore, the flow of the mixture is assumed to be incompressible as well, i.e.,
    
    \begin{equation}
    	\label{eq:divU}
    	\frac{\partial u_i}{\partial x_i} = 0.
    \end{equation}
    
    \subsection{Governing equation for $\alpha$}
    
    The governing equation for $\alpha$ could be derived from the continuity equation of the mixture system
    
    \begin{equation}
    	\label{eq:continuity}
    	\frac{\partial \rho}{\partial t} + \frac{\partial (\rho u_i)}{\partial x_i} = 0.
    \end{equation}
    Plugging in the mixture density relation, Eq. \eqref{eq:average density}, and using the divergence-free condition, Eq. \eqref{eq:divU}, we could get
    \begin{equation}
    \label{eq:alpha_0}
    \frac{\partial \alpha}{\partial t} + \frac{\partial (\alpha u_i)}{\partial x_i} = 0,
    \end{equation}
    or its equivalent form
    \begin{equation}
    \label{eq:alpha_1}
    \frac{\partial \alpha}{\partial t} + u_i \frac{\partial \alpha }{\partial x_i} = 0.
    \end{equation}
    
    Since the governing equation for $\alpha$ is directly derived from the continuity equation, it is simply another form of the equation of mass conservation. Therefore, the mass conservation is automatically satisfied in the VOF framework, making it favorable for two-phase flow simulations.
    
    \subsection{Momentum equation}
    
    One important feature for the two-phase system is the existence of phase-phase interface. Therefore, the surface tension force, which is caused by the presence of the interface, should be considered in the momentum balance of the system. Since all the conservation equations are derived using the concept of control volume, the surface tension force should be converted to a volumetric force, $\vec{F_{st}}$, before it is included in the momentum equation. Interested readers are referred to \citet{Popinet2018} for various ways of calculating $\vec{F_{st}}$. We use $\vec{F}$ to denote the sum of $\vec{F_{st}}$ and all the other body forces, if any. Then the following momentum equation is used to describe the momentum balance in the VOF framework:
    \begin{equation}
    	\label{eq:momentum}
    	\frac{\partial (\rho u_i)}{\partial t} + \frac{\partial (\rho u_i u_j)}{\partial x_j} = -\frac{\partial p}{\partial x_i} + \frac{\partial \tau _{ij}}{\partial x_j} + F_i,
    \end{equation}
    where $\tau _{ij}$ denotes the molecular shear stress tenor.
    
    \section{Reynolds and Favre decomposition}
    
    The RANS equations are derived based on Reynolds or Favre decomposition. In Reynolds decomposition, a variable, $f$, is decomposed into its mean value, $\overline{f}$, and the fluctuation component, $f'$, as given by the following:
    
    \begin{equation}
   	\label{eq:Reynolds}
   	f = \overline{f} + f',
    \end{equation}
    where the ``$\overline{\phantom{x}}$'' operator denotes time- or ensemble-averaging. One important feature of Reynolds decomposition is that the mean of the fluctuation component is zero:
    \begin{equation}
    \label{eq:Reynolds_fluctuation}
    \overline{f'} = 0.
    \end{equation}

    The Favre decomposition is based on the Favre-averaging approach, where the density-weighted mean value, $\tilde{f}$, is defined by
    
    \begin{equation}
    \label{eq:Favre_avragre}
    \tilde{f} = \frac{\overline{\rho f}}{\overline{\rho}}.
    \end{equation}
    Then the fluctuation component, $f''$, could be obtained through the following relation:
    \begin{equation}
    \label{eq:Favre}
    f = \tilde{f} + f''.
    \end{equation}    
    It should be noted that, for variable-density flows, the fluctuation component does not vanish when it is averaged, since
    \begin{equation}
    \label{eq:Favre_fluctuation}
    \overline{f''} = \overline{f}-\frac{\overline{\rho f}}{\overline{\rho}} \neq 0.
    \end{equation}

    \section{VOF with RANS modeling}
    
    \subsection{Reynolds decomposition vs. Favre decomposition}    

    Reynolds decomposition is used in strict incompressible flows where the density is constant. However, for variable-density flows, Reynolds decomposition complicates the averaged equation due to the variation in density. Take the governing equation for $\alpha$ (Eq. \eqref{eq:alpha_0}) for instance, if Reynolds decomposition is applied to $\alpha$ and $u_i$, we would end up with the following equation after averaging:
    \begin{equation}
    \label{eq:alpha_Reynolds}
    \frac{\partial \overline{\alpha}}{\partial t} + \frac{\partial (\overline{\alpha} \, \overline{u_i})}{\partial x_i} + \frac{\partial \overline{\alpha' u_i'}}{\partial x_i}= 0,
    \end{equation}
    
    The extra term, $\overline{\alpha ' u_i '}$ might be defined as the turbulent phase flux. This term could be modeled by the classic gradient-diffusion approach, i.e. by introducing the turbulent Schmidt number for $\alpha$, $\sigma_\alpha$, we could get
    
    \begin{equation}
    	\label{eq:phase_flux}
    	\overline{\alpha ' u_i '} = - \frac{\nu_t}{\sigma_\alpha}\frac{\partial \overline{\alpha}}{\partial x_i},
    \end{equation}
    where $\nu_t$ is the turbulent viscosity. Alternatively, \citet{Shirani2006} proposed
    \begin{equation}
    \label{eq:phase_flux1}
    \overline{\alpha ' u_i '} = - \frac{\nu_t}{\sigma_\alpha} \frac{\partial \overline{\alpha}}{\partial x_i} - \frac{\overline{\alpha}}{2\sigma_\alpha}\frac{\partial \nu_t}{\partial x_i},
    \end{equation}
    and \citet{Shirani2011} proposed
    \begin{equation}
    \label{eq:phase_flux2}
    \overline{\alpha ' u_i '} = - \frac{\nu_t}{\sigma_\alpha} (1-\overline{\alpha})\frac{\partial \overline{\alpha}}{\partial x_i}.
    \end{equation}        
    In all these treatments, $\sigma_\alpha$ should be calibrated against high-fidelity data.    

	In addition, there will be more additional terms in the Reynolds-averaged momentum equation for variable-density flows. Therefore, more closure models are needed for these extra terms if Reynolds decomposition is used.

    The Favre-averaging approach solves this issue by its definition. Among three forms of continuity/$\alpha$ equations, Eq. \eqref{eq:continuity} is selected for averaging since the Favre-averaged approach needs to work with density. The averaged continuity equation reads
    \begin{equation}
    \label{eq:continuity_Favre}
    \frac{\partial \overline{\rho}}{\partial t} + \frac{\partial (\overline{\rho} \tilde{u_i})}{\partial x_i} = 0.
    \end{equation}   
    The averaged momentum equation reads 
    \begin{equation}
    \label{eq:momentum_Favre}
    \frac{\partial (\overline{\rho} \tilde{u_i})}{\partial t} + \frac{\partial (\overline{\rho} \tilde{u_i} \tilde{u_j})}{\partial x_j} = -\frac{\partial \overline{p}}{\partial x_i} + \frac{\partial(\overline{\tau _{ij}} - \overline{\rho u_i '' u_j ''})}{\partial x_j} + \overline{F_i},
    \end{equation}
    where $- \overline{\rho u_i '' u_j ''}$ is the turbulent stress, and will be modeled by additional equation(s).
    
    As a matter of fact, the Favre-averaging approach is always used in widespread commercial and open-source CFD codes for variable-density flows.

    \subsection{Exclusive closure requirement for RANS-VOF}
    
    Most of the commonly encountered variable-density flows are compressible flows, where Eq. \eqref{eq:divU} does not hold. In such cases, the flow system is solved by introducing additional equation(s) of state. However, the flow of the isothermal mixture in the VOF framework belongs to a special category, i.e., variable-density incompressible flow \citep{Fan2019c}. Then an issue arises when the divergence-free condition is Favre-averaged:
    \begin{equation}
    	\label{eq:divU_Favre_full}
    	\frac{\partial \tilde{u_i}}{\partial x_i} + \frac{\partial \overline{u_i ''}}{\partial x_i} = 0.
    \end{equation}    
    According to Eq. \eqref{eq:Favre_fluctuation}, $\overline{u_i ''}$ is not zero. Therefore, the Favre-averaged velocity is not guaranteed to satisfy the divergence-free condition:
    \begin{equation}
    \label{eq:divU_Favre}
    \frac{\partial \tilde{u_i}}{\partial x_i} = 0.
    \end{equation}             
    However, without Eq. \eqref{eq:divU_Favre}, the system is not closed since there are five variables ($\overline{\alpha}$, $\tilde{u}$, $\tilde{v}$, $\tilde{w}$, $\overline{p}$) and we only have four equations (Eq. \eqref{eq:continuity_Favre} and Eq. \eqref{eq:momentum_Favre}). For the isothermal system, there is no equation of state that we could use to close the system. Therefore, enforcing the divergence-free condition for the Favre-averaged velocity (Eq. \eqref{eq:divU_Favre}) seems to be the most reasonable assumption. Before further derivations, we have to dig deeper into this assumption. Using Reynolds decomposition for both $\rho$ and $u_i$ we could get
    \begin{equation}
    \label{eq:U_Favre_Reynolds}
    \tilde{u_i}=\frac{1}{\overline{\rho}}(\overline{\rho} \, \overline{u_i}+\overline{\rho} u_i ' + \rho ' \overline{u_i} +\rho '  u_i ').
    \end{equation} 
    Plugging Eq. \eqref{eq:U_Favre_Reynolds} into Eq. \eqref{eq:divU_Favre}, we will get the following equation:
    \begin{equation}
    	\label{eq:phase_flux_Favre}
    	\frac{\partial}{\partial x_i}(\frac{\overline{\rho ' u_i '}}{\overline{\rho}})=0.
    \end{equation}
    The physical interpretation of this assumption is that, instead of the treatments given by Eqs. \eqref{eq:phase_flux}-\eqref{eq:phase_flux2}, Eq. \eqref{eq:phase_flux_Favre} shows an implicit model for the turbulent phase flux.
    
    Aside from making the system closed, Eq. \eqref{eq:divU_Favre} also allows us to explicitly formulate a governing equation for $\overline{\alpha}$. For Eq. \eqref{eq:continuity_Favre}, we could substitute $\overline{\rho}$ with the averaged version of Eq. \eqref{eq:average density}
	\begin{equation}
	\label{eq:time average density}
	\overline{\rho} = \overline{\alpha} \rho _1+(1-\overline{\alpha})\rho _2,
	\end{equation}    
    and get
    \begin{equation}
    	\label{eq:alpha_Fravre_0}
    	\frac{\partial \overline{\alpha}}{\partial t} + \frac{\partial}{\partial x_i}(\overline{\alpha} \tilde{u_i}) + \frac{\rho_2}{\rho_1 -\rho_2}\frac{\partial \tilde{u_i}}{\partial x_i} =0.
    \end{equation}
    Obviously, the last term on the l.h.s., $\frac{\rho_2}{\rho_1 -\rho_2}\frac{\partial \tilde{u_i}}{\partial x_i}$, vanishes according to Eq. \eqref{eq:divU_Favre}, and we finally get
    \begin{equation}
    \label{eq:alpha_Fravre}
    \frac{\partial \overline{\alpha}}{\partial t} + \frac{\partial}{\partial x_i}(\overline{\alpha} \tilde{u_i}) =0.
    \end{equation}    
    The importance of obtaining this formulation is that it has exactly the same form as Eq. \eqref{eq:alpha_0}. They only differ in the interpretations of involved variables. Therefore, all the sophisticated algorithms, which are designed to solve Eq. \eqref{eq:alpha_0}, are simply reused without modifications.
	
	We note that, by redefining the meaning of the ``$\overline{\phantom{x}}$'' operator from time- or ensemble-averaging to spatial filtering, one could still get Eqs. \eqref{eq:divU_Favre} and \eqref{eq:alpha_Fravre} for LES \citep{Liovic2012}. In addition to the difference in the interpretation of the ``$\overline{\phantom{x}}$'' operator and the resultant variables, there is one key difference that we must clarify. For LES, Eq. \eqref{eq:divU_Favre} is achieved by neglecting the commutation error, meaning that no extra closure is used during the derivation. Therefore, Eq. \eqref{eq:phase_flux_Favre} is the exclusive closure requirement for RANS-VOF simulations. According to our knowledge, the validity of Eq. \eqref{eq:divU_Favre} or Eq. \eqref{eq:phase_flux_Favre} could only be evaluated by high-fidelity DNS-VOF simulations, which are fortunately emerging \citep{Ling2019}.
	
	\section{Conclusions and outlooks}
	
	We reveal the very basic closure assumption that has been implicitly adopted in common RANS-VOF simulations that the Favre-averaged velocity must be divergence-free. This assumption is exclusive to RANS-VOF simulations since the flow of the isothermal mixture is variable-density incompressible. Considering the increasing popularity of RANS-VOF simulations, the validity of this closure assumption should be carefully investigated in future studies.
	
	We do not deal with any specific RANS turbulence models, since such discussion is out of the scope of the present study. However, we would like to mention that mature RANS turbulence models for VOF are still absent. Further investigations are needed for such models.
    
	\section*{Conflicts of interest}
	
	None.
	
	\section*{Acknowledgement}
	Wenyuan Fan is grateful for the support of China Scholarship Council (CSC, grant no. 201600160035).

	\bibliography{bib}

\begin{thebibliography}{11}
\providecommand{\natexlab}[1]{#1}
\providecommand{\url}[1]{\texttt{#1}}
\expandafter\ifx\csname urlstyle\endcsname\relax
  \providecommand{\doi}[1]{doi: #1}\else
  \providecommand{\doi}{doi: \begingroup \urlstyle{rm}\Url}\fi

\bibitem[Fan and Anglart(2019)]{Fan2019c}
W.~Fan and H.~Anglart.
\newblock {varRhoTurbVOF: A new set of volume of fluid solvers for turbulent
  isothermal multiphase flows in OpenFOAM}.
\newblock \emph{Computer Physics Communications}, page 106876, 2019.
\newblock \doi{10.1016/j.cpc.2019.106876}.

\bibitem[Fan et~al.(2019)Fan, Li, and Anglart]{Fan2018j}
W.~Fan, H.~Li, and H.~Anglart.
\newblock {Numerical investigation of spatial and temporal structure of annular
  flow with disturbance waves}.
\newblock \emph{International Journal of Multiphase Flow}, 110:\penalty0
  256--272, 2019.
\newblock \doi{10.1016/j.ijmultiphaseflow.2018.10.003}.

\bibitem[Hirt and Nichols(1981)]{Hirt1981}
C.~Hirt and B.~Nichols.
\newblock {Volume of fluid (VOF) method for the dynamics of free boundaries}.
\newblock \emph{Journal of Computational Physics}, 39\penalty0 (1):\penalty0
  201--225, jan 1981.
\newblock \doi{10.1016/0021-9991(81)90145-5}.

\bibitem[Ling et~al.(2017)Ling, Fuster, Zaleski, and Tryggvason]{Ling2017}
Y.~Ling, D.~Fuster, S.~Zaleski, and G.~Tryggvason.
\newblock {Spray formation in a quasiplanar gas-liquid mixing layer at moderate
  density ratios: A numerical closeup}.
\newblock \emph{Physical Review Fluids}, 2\penalty0 (1):\penalty0 1--17, 2017.
\newblock \doi{10.1103/PhysRevFluids.2.014005}.

\bibitem[Ling et~al.(2019)Ling, Fuster, Tryggvason, and Zaleski]{Ling2019}
Y.~Ling, D.~Fuster, G.~Tryggvason, and S.~Zaleski.
\newblock {A two-phase mixing layer between parallel gas and liquid streams:
  Multiphase turbulence statistics and influence of interfacial instability}.
\newblock \emph{Journal of Fluid Mechanics}, 859:\penalty0 268--307, 2019.
\newblock \doi{10.1017/jfm.2018.825}.

\bibitem[Liovic and Lakehal(2007)]{Liovic2007}
P.~Liovic and D.~Lakehal.
\newblock {Multi-physics treatment in the vicinity of arbitrarily deformable
  gas-liquid interfaces}.
\newblock \emph{Journal of Computational Physics}, 222\penalty0 (2):\penalty0
  504--535, 2007.
\newblock \doi{10.1016/j.jcp.2006.07.030}.

\bibitem[Liovic and Lakehal(2012)]{Liovic2012}
P.~Liovic and D.~Lakehal.
\newblock {Subgrid-scale modelling of surface tension within interface
  tracking-based Large Eddy and Interface Simulation of 3D interfacial flows}.
\newblock \emph{Computers and Fluids}, 63:\penalty0 27--46, 2012.
\newblock \doi{10.1016/j.compfluid.2012.03.019}.

\bibitem[Popinet(2018)]{Popinet2018}
S.~Popinet.
\newblock {Numerical Models of Surface Tension}.
\newblock \emph{Annual Review of Fluid Mechanics}, 50\penalty0 (1):\penalty0
  49--75, 2018.
\newblock \doi{10.1146/annurev-fluid-122316-045034}.

\bibitem[Saxena and Prasser(2019)]{Saxena2019}
A.~Saxena and H.-M. Prasser.
\newblock {A STUDY OF TWO-PHASE ANNULAR FLOW USING UNSTEADY NUMERICAL
  COMPUTATIONS}.
\newblock \emph{International Journal of Multiphase Flow}, 2019.
\newblock \doi{10.1016/j.ijmultiphaseflow.2019.05.003}.

\bibitem[Shirani et~al.(2006)Shirani, Jafari, and Ashgriz]{Shirani2006}
E.~Shirani, A.~Jafari, and N.~Ashgriz.
\newblock {Turbulence models for flows with free surfaces and interfaces}.
\newblock \emph{AIAA Journal}, 44\penalty0 (7):\penalty0 1454--1462, 2006.
\newblock \doi{10.2514/1.16647}.

\bibitem[Shirani et~al.(2011)Shirani, Ghadiri, and Ahmadi]{Shirani2011}
E.~Shirani, F.~Ghadiri, and A.~Ahmadi.
\newblock {Modeling and simulation of interfacial turbulent flows}.
\newblock \emph{Journal of Applied Fluid Mechanics}, 4\penalty0 (2):\penalty0
  43--49, 2011.

\end{thebibliography}
	
\end{document}